# Increased endurance of nonvolatile photonics enabled by nanostructured phase-change materials

*Jayita Dutta[*,1], Andrew Tang[1], Brian Mills[2,3], Rui Chen[1], Arnab Manna[4], Gokul Nath SJ[1], Virat Tara[1], Dennis Callahan[3], Cosmin Constantin Popescu[2], Juejun Hu[2], and Arka Majumdar[*,1,4]*

[1] Department of Electrical and Computer Engineering, University of Washington, Seattle, WA, 98195, USA.

[2] Department of Materials Science and Engineering, Massachusetts Institute of Technology, Cambridge, MA 02139, USA

[3] The Charles Stark Draper Laboratory, Cambridge, MA, 02139, USA.

[4] Department of Physics, University of Washington, Seattle, WA, 98195, USA.

E-mail: jayitad@uw.edu ; arka@uw.edu

**Funding Sources**

The research is funded by the NSF-FuSe Award, UPWARDS-UW Fellowship and NASA and DHS STTR grants. Additionally, this material is based upon work supported by the Defense Health Agency (DHA) Small Business Innovation Research (SBIR)/Small Business Technology Transfer (STTR) Programs/U.S. Army Research Institute of Environmental Medicine (USARIEM) under U.S. Army Medical Research Acquisition Activity (USAMRAA) Contract No HT942524C0068, and NSF under Award 2329088. Any opinions, findings and conclusions or recommendations expressed in this material are those of the author(s) and do not necessarily reflect the views of the Department of Defense, DHA, SBIR/STTR Programs, USARIEM, or USAMRAA.

**Abstract**

The rapid rise of artificial intelligence, and in-memory computing has reinvigorated research on scalable, energy-efficient, and reconfigurable photonic hardware. Non-volatile phase-change materials (PCMs) are attractive, as they offer large refractive index contrast, wavelength-scale footprints, and zero static power consumption. However, current PCM-based electrically controlled photonic devices are plagued by high insertion loss and low endurance. One prevalent hypothesis for these material limitations come from electromagnetic scattering in the interface and large programming volumes, respectively. Here, we validate this hypothesis by showing that nano-structuring of PCM minimizes optical loss and enhances the endurance. By tapering both ends of a wide bandgap PCM $Sb_2Se_3$ segment on a silicon waveguide, we suppressed the insertion loss by ~94% (resulting in a loss of ~0.1 dB per π phase shift). Through combining tapering and segmentation, we achieved high optical modulation amplitude (~70%), low loss (~0.5 dB per π phase shift), low-voltage (< 5V) actuation, and record high endurance greater than 100 million cycles. This work showcases the substantial advantage of nanopatterning PCMs to attain low loss and high cyclability.

## Introduction

The exponential growth of artificial intelligence (AI), and in-memory computing is reshaping modern computation, placing unprecedented demands on hardware efficiency, bandwidth, and scalability[1–6]. As AI models scale to billions of parameters and data-intensive workloads spread from cloud to edge, performance is increasingly limited by data movement[7–19]. This has led to the memory wall—where processor speed outpaces memory bandwidth—a major barrier to throughput and energy efficiency[20,21]. Conventional copper interconnects further exacerbate the problem through resistive losses and limited bandwidth[22–26], underscoring the need for new architectures like optical interconnects to overcome these bottlenecks[1,27–30].

Photonic integrated circuits (PICs), which exploit the inherent advantages of light—ultra-high bandwidth, low latency, and negligible propagation loss—enable efficient, high-throughput data transport [31–39]. While PICs have revolutionized communication, realizing their full potential for workloads like programming, neuromorphic computing, and in-memory processing requires programmable and reconfigurable components[29,40–46]. This need has catalyzed the emergence of non-volatile programmable PICs, embedding memory-like functionality directly into photonic hardware, for scalable optical computing and signal processing[46–52].

Chalcogenide-based PCMs such as germanium–antimony–telluride ($Ge_2Sb_2Te_5$, GST), $Ge_2Sb_2Se_4Te_1$ (GSST), antimony sulfide ($Sb_2S_3$, $Sb_2S_3$), and antimony selenide ($Sb_2Se_3$, $Sb_2Se_3$)[41,47,48] enable large, non-volatile, reversible refractive index changes ($\Delta n$) between micro-structural amorphous and crystalline states, supporting strong phase and amplitude modulation in micron-scale footprints[48,53–55]. They can be switched optically (ps–μs laser pulses) or electrically

(integrated microheaters), with non-volatile states that persist without static power consumption[56–60] —unlike conventional tuning mechanisms like thermo-optic effect, free-carrier dispersion, or the Pockels effect, that require continuous biasing[61–66]. Combined with broadband response and CMOS compatibility, PCMs underpin reconfigurable platforms such as programmable photonic gate arrays[35]. Yet, PCM-based photonics still faces critical bottlenecks[35,48,67]. Despite its high refractive index contrast (Δn ~2.7) and fast programming, telluride-based PCMs exemplified by GST[35,59,68,69] suffers from strong crystalline absorption (>~ 2.5 - 3 dB per $\pi$ phase shift) due to a narrow bandgap ($E_g$ ~ 0.5–0.7 eV), degrading signal quality and limiting cascading[57,70]. Endurance is also restricted (~$10^5$ cycles[71]) due to programming-induced delamination, dewetting, and elemental segregation[48,72,73], while continuous PCM films are more prone to these failure mechanisms[35,67,74,75].

In response, recent research has pursued low-loss, wide-bandgap PCMs and device-level innovations. $Sb_2S_3$ ($E_g$ ~ 2.0 eV), offers less than 1dB loss per $\pi$ phase shift; negligible crystalline-state absorption but smaller Δn (0.4–0.5) and higher programming energy[29,48,71,74,75,76]. In contrast, $Sb_2Se_3$ ($E_g$ ~ 1.2 eV) combines near-zero loss, relatively high Δn (~0.7–1), and good thermal stability, balancing loss, efficiency and scalability[31,69,76,78–80]. At the device level, patterned PCM patches and advanced heaters (e.g., graphene and Si PIN) have lowered programming energies to the nJ regime[46,48,71,72,76,81,82]. Recent work has also introduced segmented $Sb_2Se_3$ structures to enhance cycling endurance in non-volatile photonic switches, demonstrating stable operation up to ~$10^4$ cycles[78]. Subwavelength PCM structuring has also been explored to reduce insertion loss and improve switching contrast in microring-based platforms[78,82]. System-level demonstrations span reconfigurable routing, beam steering, and neuromorphic processing, and field-programmable microring WDM transceivers leveraging monolithically integrated phase-change materials[83], while platform-level integration on 300-mm silicon photonics and back-end-of-line (BEOL) deposition further improves manufacturability[35,67,77,77,84,85]. Yet no architecture simultaneously achieves ultra-low loss, high endurance, and strong contrast—key metrics for scaling PCM–PICs into high-performance programmable photonic systems.

In this work, we investigate nano-patterned $Sb_2Se_3$-clad silicon microring resonators with optimized PIN heaters for programmable photonics at telecommunication-band (1.5–1.6 µm). We hypothesize that uniform, laterally tapered films of PCM will enable low loss (quantified by the PCM-induced insertion loss per π phase shift) by circumventing modal mismatch in a waveguide and segmented islands can increase the endurance due to reduced thermal, mechanical and chemical inhomogeneity. We experimentally validated our hypothesis and demonstrated a reduced insertion loss (~0.5 dB per π phase shift) while distributing thermal load across smaller PCM segments to simultaneously achieve high endurance (>$10^8$ cycles), strong optical modulation amplitude (~70%), and small programming energies of 33 nJ (amorphization) and 25 µJ (crystallization). By resolving the loss–endurance–contrast inter-dependence, this architecture advances PCM–PIC scalability for programmable photonics.

## Results and Discussion

We first independently evaluate both experimentally and in simulation, two distinct PCM–waveguide integration strategies—(i) laterally (i.e., in-plane) tapered continuous thin film, and (ii) segmented islands—using $Sb_2Se_3$ integrated on silicon microring resonators. Then we combine both patterns to create tapered $Sb_2Se_3$ segments. We actuate the phase transition using PIN microheater geometry (**Figure 1a**). The heater consists of a 1 µm-wide intrinsic silicon (i-Si) region, with lateral separations of 0.2 µm and 0.3 µm from the partially etched waveguide edge to the p++ and n++ doped regions, respectively. This asymmetric spacing of the doping regions is intentional, concentrating carrier-induced heating near the high-field region while keeping the more loss-inducing n-type carriers farther away, thereby enhancing heating efficiency without sacrificing optical quality factor. The same heater length was used for all devices, and the heater was implemented as a continuous (non-segmented) structure to ensure uniform thermal distribution across the PCM region (details in supplementary section S3 and Figure S7).

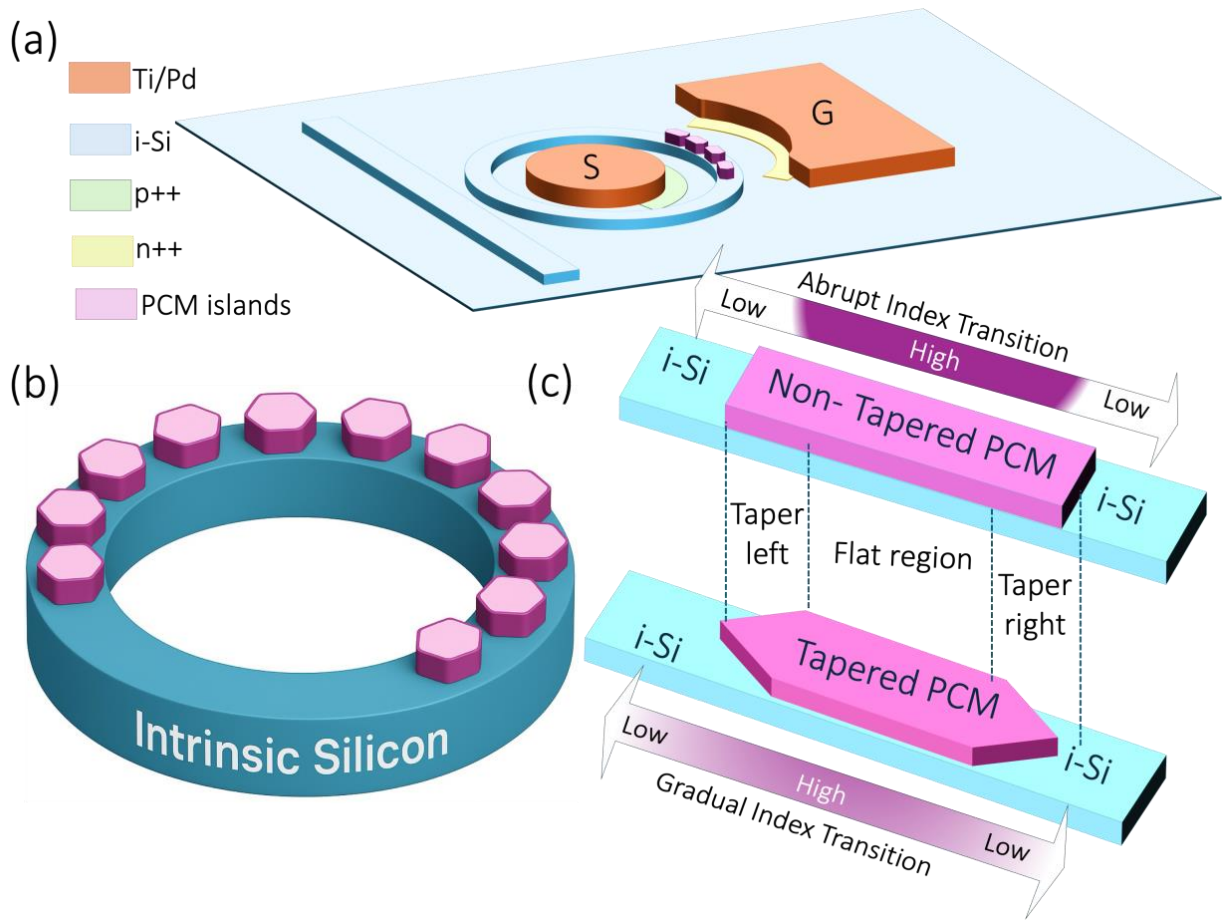

**Figure 1: A silicon microring resonator loaded with 20 nm thick and 450 nm wide $Sb_2Se_3$ layer, patterned into discrete islands. (a).** Schematic of silicon ring resonator loaded with tapered $Sb_2Se_3$ islands and actuated by optimized silicon PIN heater for low power, low loss and high endurance programming. **(b).** Zoom-in view of intrinsic silicon ring integrated with tapered $Sb_2Se_3$ islands. **(c)** Zoom-in view of a

Sb$_2$Se$_3$ island with and without tapering. The non-tapered PCM has abrupt interfaces that cause sudden refractive-index discontinuities, while the tapered PCM introduces a gradual index transition from silicon to Si– Sb$_2$Se$_3$ composite and back, reducing modal overlap and interfacial scattering losses. $L_\text{total} = L_c + L_t$ $L_{total} = L_c + L_t$, where $L_{total}$ is the total length of the PCM on the silicon waveguide, $L_c$ and $L_t$ are the length of the flat PCM region and the tapers, respectively. For non-tapered islands $L_t = 0$ and $L_{total} = L_c$.

In the segmented PCM designs (**Figure 1b and c**), the Sb$_2$Se$_3$ layer was patterned into a series of discrete islands placed along the microring. Each island comprises a central flat region of Sb$_2$Se$_3$ flanked by laterally tapered ends that provide a smooth transition between the PCM-loaded and the bare silicon waveguide, thereby minimizing scattering losses at the interfaces. **Figure 1b** shows the 3D view of the intrinsic silicon ring integrated with these tapered Sb$_2$Se$_3$ islands, while **Figure 1c** presents a zoomed-in view of a single island highlighting the taper and non-tapered sections. For a fair comparison across devices, the total PCM coverage length ($L_{total}$) is kept the same as 10 μm. For devices with multiple islands, the effective length per island ($L_i$) is given by Equation (1).

$$L_i = \frac{L_{total}}{N} \tag{1}$$

where $N$ is the number of islands. Each island further consists of a central PCM section of length $L_c$ and two tapered ends with a combined length $L_t$. Thus, for tapered island designs, the total island length is given by Equation (2). The corresponding non-tapered island configuration, consisting of PCM patches with abrupt interfaces to the silicon waveguide, is shown in Section 1 of Supplementary Information (Figure S1).

$$L_i = L_c + L_t = \frac{L_{total}}{N} \tag{2}$$

**Figure 2** summarizes the structural characterization of the fabricated Sb$_2$Se$_3$-integrated microring resonators. The optical micrograph (**Figure 2a**) shows the integrated PIN microheater and patterned PCM region, confirming uniform device fabrication. The scanning electron microscope (SEM) image (**Figure 2b**) highlights precise alignment of $Sb_2Se_3$ islands over the partially etched Si waveguide with smooth island edges and clean lift-off, indicative of high-quality patterning. The atomic force microscopy (AFM) scans of laterally tapered multiple-island

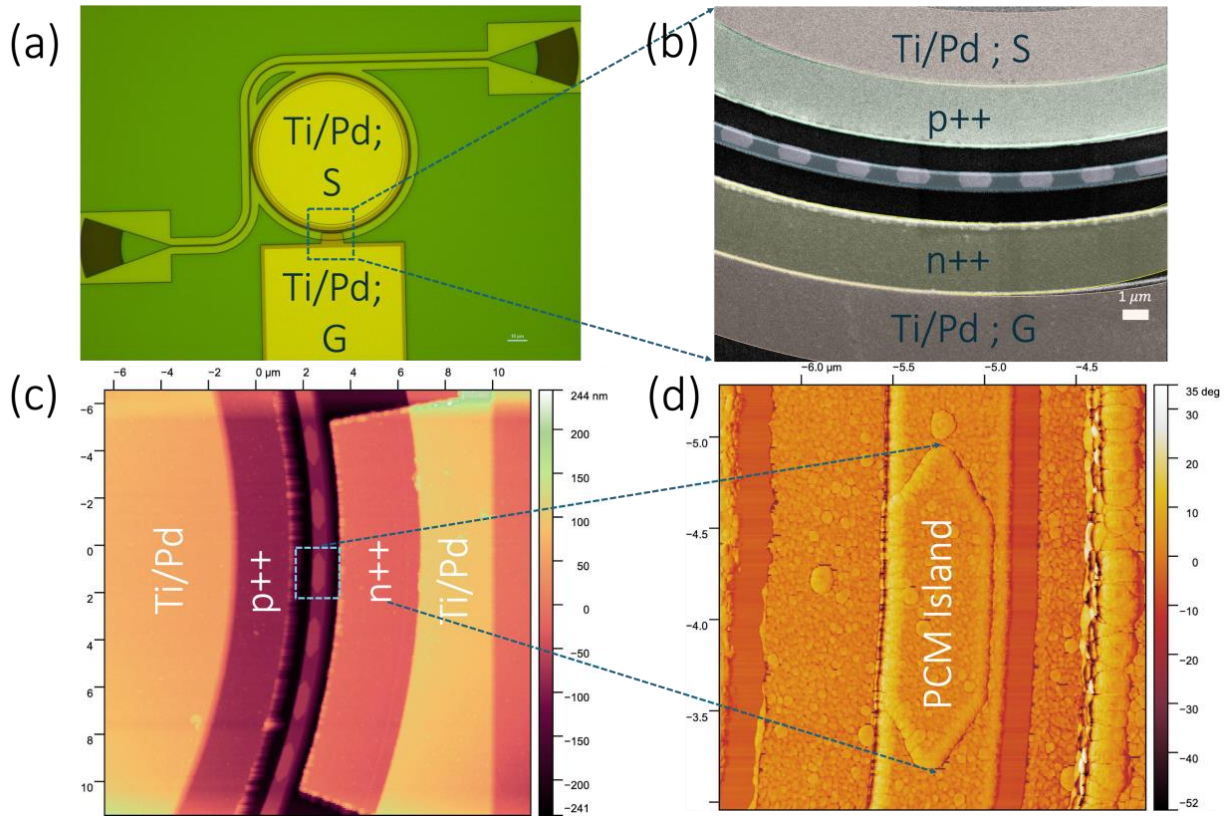

PCM structures. The higher-magnification AFM scan (**Figure 2c-d**) further confirm the designed taper geometry, 1-μm inter-island spacing, and uniform PCM thickness of ~20 ± 1 nm with minimal (< 1 nm) variation, demonstrating excellent deposition uniformity and strong $Sb_2Se_3$–Si adhesion (See Supplementary Figure S12).

**Figure 2: Images of the $Sb_2Se_3$ integrated silicon ring resonator. (a).** Optical microscope image of a silicon microring, showing the microring, grating couplers, and electrical contacts. **(b).** Scanning electron

microscope (SEM) image providing a zoomed-in view of the heater section, highlighting the precise alignment of $Sb_2Se_3$ islands atop the Si waveguide within the PIN region. The $Sb_2Se_3$ thin film, n++, and p++ doped silicon regions are represented by false colors (violet, yellow, and green respectively). **(c)**. Atomic force microscope (AFM) topography image showing the array of $Sb_2Se_3$ islands with lateral tapering on the silicon ring, confirming uniform deposition and well-defined taper geometry. **(d).** Zoomed-in AFM image of a single tapered $Sb_2Se_3$ island, illustrating the smooth transition profile introduced by the taper design.

Additional microscope, SEM and AFM characterizations of devices with different island numbers and geometries (Supplementary Figures S9–S13) validate consistent morphology and alignment across all integration schemes. We restrict our quantitative analysis to 11 islands to ensure consistency among the geometry. Due to the fixed total length of 10 μm, more islands lead to a rhomboidal shape with zero center parts (see Supplementary Figure S14 for 13 islands).

## 2.1 Effect of tapered PCM ends

The schematic (**Figure 3b**, inset) illustrates the taper expanding from near-zero width to ~450 nm while maintaining a 20 nm thickness. This gradual index transition reduces modal overlap with $Sb_2Se_3$ (**Figure 1c**), thereby minimizing interfacial scattering and slightly lowering phase modulation[86–89].

**Figure 3a** shows averaged transmission spectra over 10 programming cycles of a device with taper length $L_t$ = 0.8 μm for amorphous (a-$Sb_2Se_3$) and crystalline (c-$Sb_2Se_3$) states. Electrical programming used optimized pulses: 1.5 ms at 1.28 V for crystallization and 400 ns at 2.83 V for amorphization, producing a resonance shift $Δλ ≈ 0.25$ nm. The programming voltages were determined by gradually increasing the applied bias from slightly above the diode turn-on voltage (~0.8 V) while monitoring the resonance spectra and selecting the minimum voltage at which $Δλ$ reached a repeatable saturation value. These saturation-based criteria ensured complete crystallization and amorphization for each geometry and avoided partial phase transitions.

It is important to note that after the initial few thousand cycles, crystallization could be achieved using much shorter microsecond pulses (~80 µs) at the same applied voltage. Periodic wavelength sweeps confirmed that the resonance shift (Δλ), phase shift, and extinction ratio remained unchanged, indicating that both pulse durations yield an identical saturated crystalline optical state. This behavior reflects an early-cycle conditioning phase—likely associated with partial structural ordering and reduced nucleation barriers—after which the device reaches a stabilized switching regime. The programming voltage remains unchanged throughout extended cycling, and the shortened pulse width enables reduced programming energy without affecting optical performance. To quantify the effect of tapering on loss and phase modulation, we extracted the optical parameters of single $Sb_2Se_3$ segments from resonance spectra (details in Section 4.3). At longer taper lengths, the $Q$ values for the crystalline and amorphous states were nearly identical ($Q_a \approx 1.54 \times 10^4$, $Q_c \approx 1.51 \times 10^4$ for $L_t = 0.8$ µm; see Supplementary Section 5.7 for PIN-integrated bare microring reference measurements), indicating strong suppression of scattering losses, while the corresponding phase shift (~0.156π) and normalized loss (0.064 dB per π phase shift) further confirm the effectiveness of long tapers in minimizing interfacial scattering.

**Figure 3b** summarizes the extracted PCM-induced loss (left axis) and phase shift (right axis) as a function of tapering length $L_t$ (0–0.8 µm), keeping the total PCM length $L_{total}$ = 10 µm constant. For each taper length, the SET and RESET voltages were independently optimized to reach a saturated and repeatable Δλ prior to extracting optical parameters, ensuring that all geometries were evaluated at their fully switched terminal states. A monotonic reduction in scattering loss—from ~1.6 dB per π phase shift (no taper) to ~0.1 dB per π phase shift ($L_t$ = 0.8

µm)—is observed, consistent with smoother index transitions at the tapered interfaces. However, since longer tapers reduce not only the effective PCM interaction length but also the total PCM area that overlaps the guided mode, the net phase shift decreases from ~0.187 π to ~0.156 π, consistent with an approximately constant phase shift per unit PCM area. The dotted lines represent linear trends, with error bars showing variations across three devices per taper length. Beyond $L_t$ = 0.8 µm, additional tapering yields minimal further loss reduction but a significant decline in phase shift (Supplementary Figure S16), emphasizing the trade-off between scattering-loss mitigation and phase-modulation efficiency.

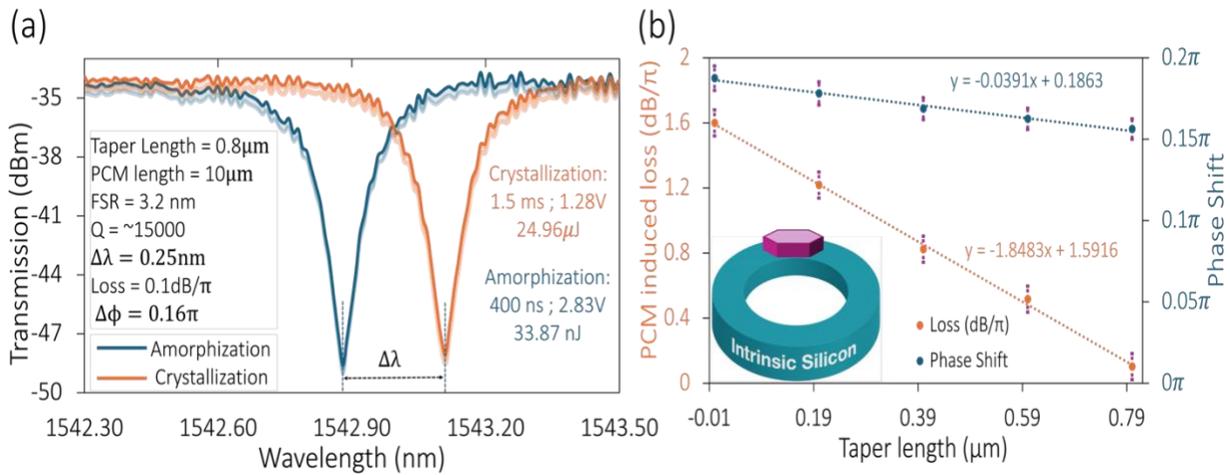

**Figure 3: Effect of tapered ends on optical loss and phase modulation in a microring resonator with a single integrated $Sb_2Se_3$ segment. (a).** Measured transmission spectra of the microring resonator in the crystalline (SET) and amorphous (RESET) states of $Sb_2Se_3$. Crystallization (orange) was achieved with a 1.5 ms pulse at 1.28 V (~13 mA; 24.96 µJ), while amorphization (blue) was induced using a 400 ns pulse at 2.83 V (~29 mA; 33.87 nJ). A resonance wavelength shift (Δλ) of ~0.25 nm between the two states corresponds to an induced phase shift (Δϕ) of ~0.16 π. Spectra are averaged over ten programming cycles; shaded regions denote standard deviation **(b).** Extracted PCM round-trip optical loss (left axis) and corresponding phase shift (right axis) as a function of tapering length for a single 10 µm-long $Sb_2Se_3$ segment. A monotonic reduction in scattering loss—from ~1.6 to ~0.1 dB per π phase shift—is observed with increasing taper length, attributed to smoother refractive-index transitions at the PCM–Si interfaces. However, the phase shift decreases from 0.187π to 0.156π due to the reduced effective PCM interaction

length. The dotted lines represent linear trend, and the inset shows a 3D schematic of the microring resonator with a single tapered $Sb_2Se_3$ segment.

## 2.2 Effect of PCM islands: Tapered and Non-Tapered

Then we explore the effect of segmentation on the device performance. We expect to achieve higher endurance due to reduced thermal, mechanical and chemical inhomogeneity in smaller PCM islands, which not only mitigates local stress accumulation and interfacial degradation but also suppresses atomic diffusion owing to the smaller volume, fewer atoms, and shorter diffusion lengths. First, we tested the segmentation with non-tapered islands on a microring resonator and then compared their performance with tapered segmentation. The islands were distributed uniformly along the ring with 1 μm gaps separating adjacent PCM segments. We investigated four configurations—1, 2, 7, and 11 islands, keeping the total PCM length fixed at 10 μm (Details in Supplementary Table S1 of Supplementary information).

**Figure 4** compares the PCM-induced excess loss per π phase shift ($\alpha'_{PCM}$) for tapered and non-tapered segments. For both cases, the loss increases monotonically with the number of islands due to the greater number of Si/PCM junctions introducing additional scattering. However, the magnitude of loss is dramatically lower in the tapered case and further validates our hypothesis that tapering reduces loss due to less modal mismatch. The device with 11 tapered $Sb_2Se_3$ islands exhibits an excess loss of only ~0.53 dB per π phase shift, compared to ~2.77 dB per π phase shift for its non-tapered equivalent. The linear trend yields a slope of ~0.11 dB per island for non-tapered devices compared to only ~0.04 dB per island for tapered devices. We note that, however, the effective optical phase shift ($\Delta\phi$) is slightly smaller for the tapered configuration (~0.15π) than for the non-tapered case (~0.19π), since tapering reduces the PCM's effective optical length (see Supplementary Section 5.4).

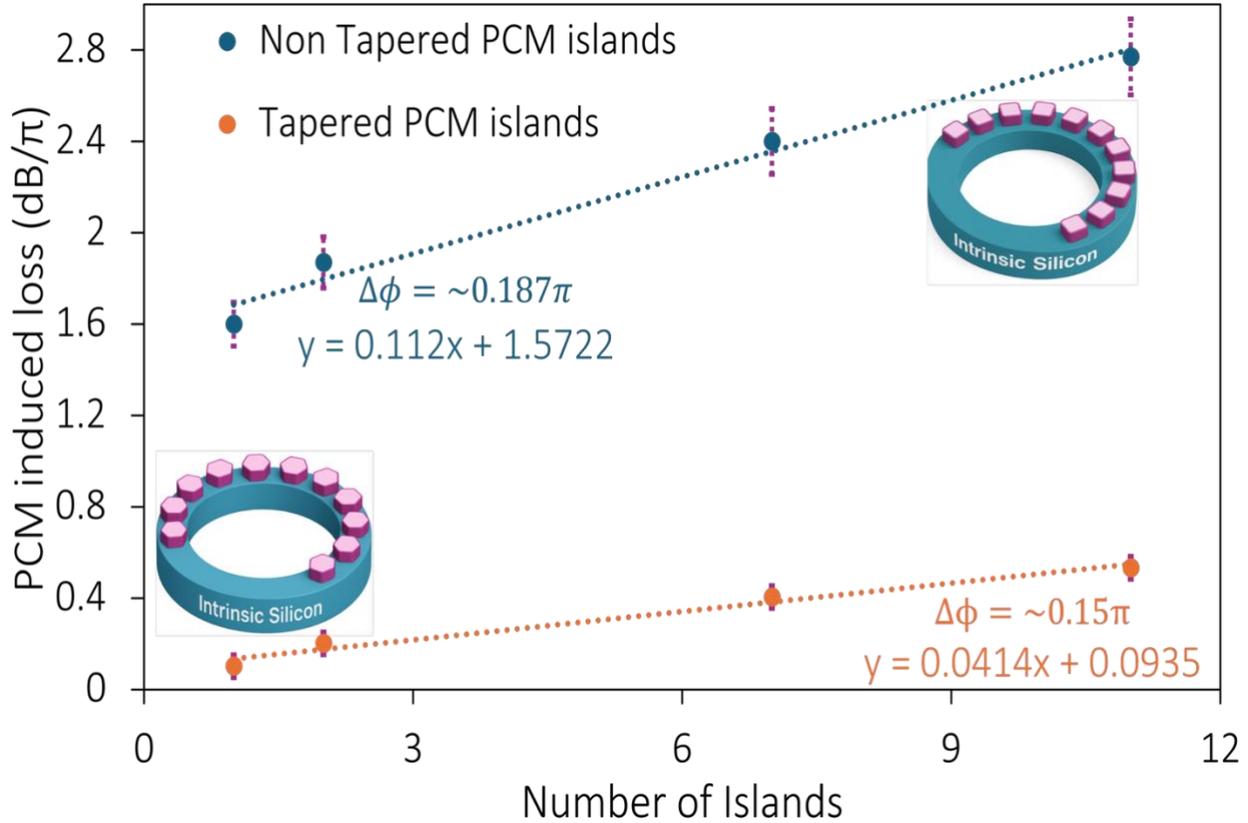

**Figure 4: Measured PCM-induced round-trip loss (in dB per π phase shift) versus number of Sb$_2$Se$_3$ islands integrated on a silicon microring resonator, for non-tapered (blue) and tapered (orange) PCM island configurations.** In both cases, the loss increases monotonically with island count due to additional Si/PCM interfaces, but the magnitude is significantly smaller for the tapered design. Linear trend yield slopes of 0.112 dB per island for non-tapered and 0.041 dB per island for tapered configurations, corresponding to a nearly three-fold reduction in interface-induced scattering. The effective optical phase shift (Δϕ) slightly decreases from ~0.187π in non-tapered to ~0.15π in tapered devices, attributed to the reduced effective PCM length in the latter. Insets illustrate the two geometries: (top) non-tapered Sb$_2$Se$_3$ islands with abrupt Si/PCM junctions and (bottom) tapered Sb$_2$Se$_3$ islands providing smooth adiabatic transitions that minimize mode mismatch.

**Figure 5** presents the endurance and contrast characteristics of the segmented PCM devices, while keeping a constant total PCM length of 10 μm. The programming conditions were first optimized for each geometry using fixed pulse durations (1.5 ms for crystallization and 400 ns for amorphization) while incrementally adjusting the applied voltage to achieve a saturated and repeatable resonance wavelength shift (Δλ). The optimized programming voltage was then kept constant throughout endurance cycling.

Minor variations were observed during the first ~10–20 cycles, attributed to structural relaxation and interface stabilization. After the initial few thousand cycles, we observed that the same saturated Δλ could be achieved with a shorter crystallization pulse (~80 μs) at the same voltage. This reduction reflects an early-cycle conditioning phase rather than material degradation. Owing to the high PCM film uniformity achieved via thermal evaporation and the thick (~100 nm) ALD $Al_2O_3$ encapsulation, stable and repeatable switching was established after this initial conditioning phase and maintained throughout extended cycling. Detailed pulse and voltage optimization procedures are provided in Supplementary Section 4.5.

In both tapered and non-tapered devices, the programming voltage and power increase with the number of islands because longer heater length is used for segmented designs to accommodate the additional spacings between PCM islands (Table S1 in Supplementary Information), enlarging the heated volume during programming. However, this voltage increase is much more pronounced in non-tapered devices, reaching ~4.58 V for amorphization in the 11-island case, compared to only ~4.1 V for the tapered counterpart (Supplementary Table S2). The lower voltage requirement in tapered devices is attributed to smoother Si– $Sb_2Se_3$ transitions that improve thermal confinement and enable more efficient heat delivery to the PCM region, allowing programming at reduced input voltage and energy.

Segmentation greatly enhances endurance as we hypothesized earlier in the paper. The improvement is evident in both geometries: endurance increases from ~3 × 10⁵ cycles for a single continuous PCM film to > 4 × 10⁷ cycles for 11 non-tapered islands and exceeds ~10⁸ cycles for 11 tapered islands, setting a record for electrically controlled optical PCM devices. The latter represents more than a 2.5-fold enhancement in endurance compared with the non-tapered case. The experiment for the device with 11 tapered islands was stopped before the device failed only due to the interest of time. The sustained optical contrast, together with the stable resonance wavelength shift (Δλ) measured at logarithmically sampled intervals, confirms reversible phase modulation during extended cycling up to $10^8$ cycles. (Figure S18-S20, Supplementary Information). The superior endurance in tapered devices possibly stems from two synergistic factors: (i) reduced stress owing to smaller active PCM volumes, which suppresses failure due to film delamination and PCM dewetting[75]; and (ii) minimized elemental segregation due to decreased temperature gradient and thermal diffusion. Post-cycling SEM characterization further demonstrates that the introduction of tapering and segmentation progressively improves structural stability, in agreement with the measured endurance scaling (Figure S15, Supplementary Information).

We also calculated the optical modulation amplitude[87,88] (i.e., the optical contrast) $\eta = \frac{T_c - T_a}{T_c + T_a}$, with $T_c$ and $T_a$ being the resonant transmission amplitudes in the crystalline and amorphous states, respectively (see Section 5.5, Supplementary Information for measurement details). The contrast ratio exhibits a slight decrease with increasing island number for both configurations, because segmentation introduces multiple Si–PCM interfaces and shorter PCM sections, which reduce the effective optical confinement within each island and increase interfacial scattering, thereby lowering the overall contrast. Nevertheless, tapered islands consistently yield a much higher contrast ratio—for example, ~70% for 11 tapered islands versus ~20% for their non-tapered counterparts— because tapering suppresses interfacial scattering at the Si/PCM junctions, thereby reducing PCM-induced excess loss and keeping the resonator closer to its critical-coupling condition. This enables higher contrast compared to the non-tapered counterpart, even though the PCM mode overlap per island is slightly smaller.

Segmenting the PCM layer into discrete islands markedly improves endurance by mitigating thermal-mechanical fatigue, and chemical inhomogeneity—thereby mitigating local stress, interfacial degradation, and elemental segregation—while tapering each island minimizes scattering losses and enhances optical modulation contrast. Although both geometries show increasing programming voltage with the number of islands, the rise is substantially smaller in the tapered design because smoother Si–$Sb_2Se_3$ transitions improve thermal confinement and heat delivery efficiency. Consequently, tapered PCM islands achieve a favorable combination of endurance (> $10^8$ cycles), contrast ratio (~70%), optical loss (~0.5 dB per $\pi$ phase shift), and programming voltage (~4 V), outperforming non-tapered devices across all metrics.

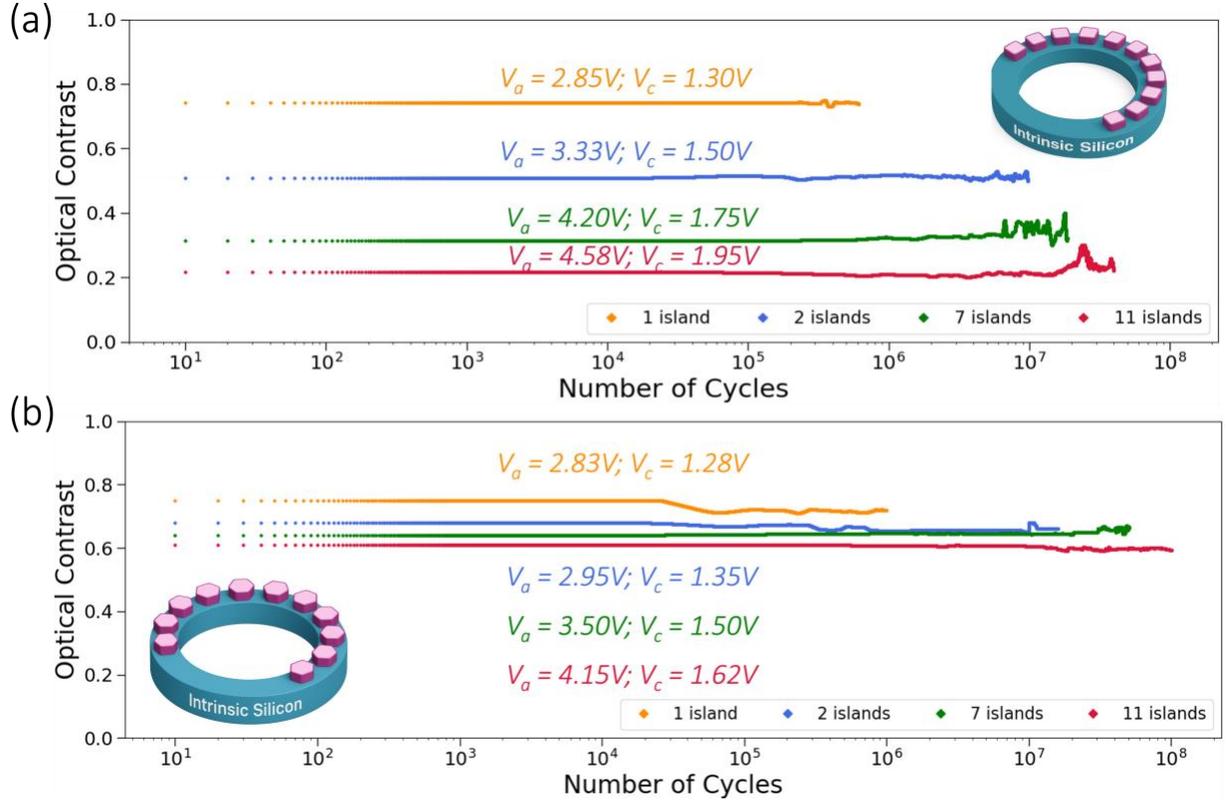

**Figure 5: Comparison of endurance and optical contrast between non-tapered and tapered $Sb_2Se_3$ island configurations integrated on silicon microring resonators.** (a). Measured contrast ratio versus programming cycles for devices with 1, 2, 7, and 11 non-tapered $Sb_2Se_3$ islands (constant total PCM length = 10 μm). Increasing the number of islands enhances endurance from ~3 × $10^5$ cycles (1 island) to > 4 × $10^7$ cycles (11 islands), accompanied by a moderate rise in programming voltage from Va = 2.85 V (1 island) to 4.58 V (11 islands) and Vc = 1.30 V (1 island) to 1.95 V (11 islands). However, the contrast ratio decreases from ~78% to ~20% as island count increases, due to additional Si/PCM interfaces introducing scattering losses. (b). Measured contrast ratio versus programming cycles for devices with 1, 2, 7, and 11 tapered $Sb_2Se_3$ islands. Tapered islands exhibit substantially higher endurance (> $10^8$ cycles for 11 islands) and improved optical contrast (~70% for 11 islands) while requiring lower programming voltages (Va = 2.83 V, Vc = 1.28 V → Va = 4.15 V, Vc = 1.62 V) than their non-tapered counterparts. Va and Vc denotes the programming voltage for amorphization and crystallization respectively. Insets show 3D schematics of the two geometries: (top) non-tapered $Sb_2Se_3$ islands with abrupt Si/PCM boundaries and (bottom) tapered $Sb_2Se_3$ islands enabling smooth adiabatic-like optical mode transitions. The optical contrast was calculated as η = (Tc − Ta)/(Tc + Ta), where Tc and Ta correspond to the steady-state transmission levels measured at the same fixed probe wavelength after crystallization (SET) and amorphization (RESET), respectively (see Supplementary Section 5.6 for absolute transmission levels).

These findings establish tapered segmentation as a robust, CMOS-compatible strategy for realizing high-endurance, low-loss, and high-contrast non-volatile photonic phase shifters.

## 3. Outlook

Looking ahead, fabrication refinements such as reducing the inter-island gap from 1 μm to ~0.2 μm—feasible via precision dry etching—could lower the overall thermal mass, thereby reducing programming voltage and power[92]. Simulations indicate that such scaling can recover part of the lost contrast ratio and further reduce the normalized round-trip loss to ~0.1 dB per π phase shift, approaching the performance of an optimized single-tapered segment while retaining the high endurance of multi-island designs (see Supplementary Section 2.2, Figures S3–S4). Furthermore, implementing fully automated endurance measurement systems would minimize alignment drift and manual intervention, enabling continuous stress testing to a billion cycles and beyond.

Beyond binary modulation, partial crystallization of $Sb_2Se_3$ enables well-defined intermediate states that can support multilevel optical tuning. While the present work focuses on binary operation to isolate endurance and loss scaling, segmentation may further improve state granularity by confining switching to smaller PCM volumes. This architecture therefore presents a promising pathway toward stable multilevel operation for high-density optical computing and neuromorphic photonic systems.

We have systematically investigated the nanopatterning technique in $Sb_2Se_3$-based nonvolatile phase shifters in terms of several key optical metrics, including insertion loss, optical modulation amplitude, phase modulation, programming voltage and endurance. These include (i) a single continuous PCM segment with tapers at both ends, (ii) segmentation into multiple islands in tapered and non-tapered forms. In the single-segment design, extending the taper length effectively suppressed junction scattering, reducing excess loss to ~0.1 dB per π phase shift with minimal optical phase-shift reduction (~15%). In the non-tapered multi-island design, we showed that segmentation enhances endurance by reducing the PCM volume switched, but at the cost of increased optical loss and reduced contrast ratio due to introduction of multiple PCM/Si junctions. By introducing tapers to each island, we achieved low optical loss and record-high endurance in the level of billion cycles. Through such nanoengineering, we establish a practical and manufacturable design paradigm for PCM-based programmable silicon photonics, bridging the gaps in endurance, insertion loss, programming voltage and scalability. Our work thus offers a compelling route toward next-generation reconfigurable photonic systems for AI acceleration, optical computing, and large-scale programmable photonic networks.

## 4. Experimental Section/Methods

### 4.1. Fabrication Methods

All device variants were fabricated on 220-nm silicon-on-insulator (SOI) chips with a 2 μm buried oxide, allowing quantitative comparison. The detailed fabrication steps are shown in Figure S6 of Supplementary Information. The 500 nm-wide bus waveguides and microrings were patterned via electron-beam lithography (EBL), followed by fluorine-based inductively coupled plasma etching, which left a 100 nm silicon slab for doping. All microrings have a bus-ring gap of 260 nm to achieve near-critical coupling. The p++ and n++ regions of the integrated microheater were defined by ion implantation, followed by Ti/Pd metallization for electrical contacts. In an optimized geometry for COMS compatible voltage, the p++ and n++ doping regions are offset from the waveguide edge by 0.2 μm and 0.3 μm, respectively, leading to an intrinsic region of ~1 µm[48]. A cross-sectional schematic of the optimized lateral PIN heater integrated with the Si waveguide is shown in Supplementary Figure S7, highlighting the asymmetric dopant placement used to balance thermal confinement with low optical loss. The PIN microheaters have a turn-on voltage (~0.8 V) and low series resistance (~50 Ω) in Supplementary Figure S8.

Sb$_2$Se$_3$ regions were introduced in a fifth EBL step using a bilayer lift-off process to define tapered and non-tapered islands. A bilayer resist stack comprising a copolymer of poly(methyl methacrylate-co-methacrylic acid) [P(MMA–MAA)] (~200 nm, written at exposure level 8 (EL8) with aperture 6 (A6, ~200 nm beam size)) and a top layer of poly(methyl methacrylate) (PMMA 950 K A4, ~200 nm) was spun sequentially to produce a controlled undercut profile, thereby improving lift-off quality and pattern reproducibility. The bi-layer lithography allowed the creation of small features, including the sharp taper ends and PCM islands with sub-100 nm edge fidelity. Additional details on Sb$_2$Se$_3$ deposition and thin-film characterization are provided in Section 3.1.4 of the Supplementary Information. The PCM width was chosen to be 50 nm smaller than the silicon waveguide to compensate for overlay tolerance and lift-off variation. Finally, the PCM was encapsulated by ~100 nm of Al$_2$O$_3$ deposited by atomic layer deposition (ALD) at 150 °C, preventing oxidation and thermal reflow—both essential for maintaining high endurance. Additional fabrication details are provided in Section 3 of the Supplementary Information.

**4.2. Optical Transmission measurements**

Optical characterization was performed using a vertical fiber-coupling setup angled at 25°, integrated with a thermoelectric controller (TEC; TE Technology TC-720) to maintain the stage temperature at 26 °C. A tunable continuous-wave laser (Santec TSL-510) provided the input light, with its polarization optimized using a manual fiber polarization controller (Thorlabs FPC526) to maximize coupling efficiency. The transmitted power was recorded using a low-noise power meter

(Keysight 81634B). For electrical programming, voltage pulses from a pulse function arbitrary generator (Keysight 81160A) were delivered to on-chip metal contacts via microprobes (Cascade Microtech DPP105-M-AI-S). All instruments—including the laser, power meter, TEC, and pulse generator—were automated through a LabVIEW interface[93]. A detailed description of the optical measurement setup can be found in Supplementary Section 3.2.

## 4.3. Extraction of Optical Parameters: Quality factor, loss and phase shift

This section outlines the procedure used to extract the quality factor ($Q$), excess round-trip loss due to the PCMs ($\alpha_{PCM}$), phase shift ($\Delta\phi$) induced by PCM, and normalized round-trip loss ($\alpha'_{PCM}$) for all device geometries discussed in the main text. The analysis was applied consistently across single-segment, non-tapered, and tapered island configurations.

The loaded quality factor ($Q$) was extracted by fitting Lorentzian to the resonance dips in the transmission spectrum, using Equation (3)[94].

$$Q = \frac{\lambda_0}{FWHM} \qquad (3)$$

where $\lambda_0$ is the resonant wavelength and the FWHM is the full width half maximum in nanometers.

Since the resonator operates near critical coupling, the excess round-trip loss due to the PCMs ($\alpha_{PCM}$) was evaluated from the $Q$-factors of the two states using Equation (4)[49]

$$\alpha_{PCM}(dB) = \frac{2\pi\lambda_0}{FSR}\left(\frac{1}{Q_c} - \frac{1}{Q_a}\right) \times \frac{10}{\ln(10)} \qquad (4)$$

where, $Q_c$ and $Q_a$ represents the Q-factor for crystalline and amorphous states, respectively.

The phase shift ($\Delta\phi$) induced by PCM was calculated using Equation (5)[48].

$$\Delta\phi = \frac{\Delta\lambda}{FSR} \times 2\pi \qquad (5)$$

To enable a fair comparison across all taper geometries, the loss was normalized to a per-π phase shift basis using Equation (6).

$$\alpha'_{PCM}(dB/\pi) = \frac{\alpha_{PCM}(dB)}{\Delta\phi/\pi} \qquad (6)$$

For example, at $L_t$ = 0.8 μm, the phase shift was ~0.156 π and the normalized loss 0.064 dB per π phase shift ($\alpha_{PCM}$ ≈ 0.01 dB), confirming that long tapers effectively minimize Si-PCM junction scattering.

## Acknowledgements

Part of this work was conducted at the Washington Nanofabrication Facility/Molecular Analysis Facility, a National Nanotechnology Coordinated Infrastructure (NNCI) site at the University of Washington. We would also like to thank Professor M.P. Anantram for the valuable discussions.

## Data Availability Statement

The data that support the findings of this study are available from the corresponding authors upon request.

## References

[1] D. V. Christensen, R. Dittmann, B. Linares-Barranco, A. Sebastian, M. Le Gallo, A. Redaelli, S. Slesazeck, T. Mikolajick, S. Spiga, S. Menzel, I. Valov, G. Milano, C. Ricciardi, S.-J. Liang, F. Miao, M. Lanza, T. J. Quill, S. T. Keene, A. Salleo, J. Grollier, D. Marković, A. Mizrahi, P. Yao, J. J. Yang, G. Indiveri, J. P. Strachan, S. Datta, E. Vianello, A. Valentian, J. Feldmann, X. Li, W. H. P. Pernice, H. Bhaskaran, S. Furber, E. Neftci, F. Scherr, W. Maass, S. Ramaswamy, J. Tapson, P. Panda, Y. Kim, G. Tanaka, S. Thorpe, C. Bartolozzi, T. A. Cleland, C. Posch, S. Liu, G. Panuccio, M. Mahmud, A. N. Mazumder, M. Hosseini, T. Mohsenin, E. Donati, S.

Tolu, R. Galeazzi, M. E. Christensen, S. Holm, D. Ielmini, N. Pryds, *Neuromorph. Comput. Eng.* **2022**, *2*, 022501.

[2] S. S. Gill, M. Xu, C. Ottaviani, P. Patros, R. Bahsoon, A. Shaghaghi, M. Golec, V. Stankovski, H. Wu, A. Abraham, M. Singh, H. Mehta, S. K. Ghosh, T. Baker, A. K. Parlikad, H. Lutfiyya, S. S. Kanhere, R. Sakellariou, S. Dustdar, O. Rana, I. Brandic, S. Uhlig, *Internet of Things* **2022**, *19*, 100514.

[3] A. H. Kelechi, M. H. Alsharif, O. J. Bameyi, P. J. Ezra, I. K. Joseph, A.-A. Atayero, Z. W. Geem, J. Hong, *Symmetry* **2020**, *12*, 1029.

[4] I. L. Markov, *Nature* **2014**, *512*, 147.

[5] R. Mayer, H.-A. Jacobsen, *ACM Comput. Surv.* **2020**, *53*, 3:1.

[6] S. Zhu, T. Yu, T. Xu, H. Chen, S. Dustdar, S. Gigan, D. Gunduz, E. Hossain, Y. Jin, F. Lin, B. Liu, Z. Wan, J. Zhang, Z. Zhao, W. Zhu, Z. Chen, T. S. Durrani, H. Wang, J. Wu, T. Zhang, Y. Pan, *Intelligent Computing* **2023**, *2*, 0006.

[7] Z. Liu, A. Ali, P. Kenesei, A. Miceli, H. Sharma, N. Schwarz, D. Trujillo, H. Yoo, R. Coffee, N. Layad, J. Thayer, R. Herbst, C. Yoon, I. Foster, in *2021 3rd Annual Workshop on Extreme-Scale Experiment-in-the-Loop Computing (XLOOP)*, **2021**, pp. 15–23.

[8] W. Woods, C. Teuscher, in *2017 IEEE/ACM International Symposium on Nanoscale Architectures (NANOARCH)*, **2017**, pp. 103–108.

[9] R. Urata, H. Liu, K. Yasumura, E. Mao, J. Berger, X. Zhou, C. Lam, R. Bannon, D. Hutchinson, D. Nelson, L. Poutievski, A. Singh, J. Ong, A. Vahdat, **2022**, DOI 10.48550/arXiv.2208.10041.

[10] H. Zhou, J. Dong, J. Cheng, W. Dong, C. Huang, Y. Shen, Q. Zhang, M. Gu, C. Qian, H. Chen, Z. Ruan, X. Zhang, *Light Sci Appl* **2022**, *11*, 30.

[11] A. Amirsoleimani, F. Alibart, V. Yon, J. Xu, M. R. Pazhouhandeh, S. Ecoffey, Y. Beilliard, R. Genov, D. Drouin, *Advanced Intelligent Systems* **2020**, *2*, 2000115.

[12] N. Maslej, L. Fattorini, E. Brynjolfsson, J. Etchemendy, K. Ligett, T. Lyons, J. Manyika, H. Ngo, J. C. Niebles, V. Parli, Y. Shoham, R. Wald, J. Clark, R. Perrault, **2023**, DOI 10.48550/arXiv.2310.03715.

[13] J. Dutta, P. Deshpande, B. Rai, *SN Appl. Sci.* **2021**, *3*, 657.

[14] J. Dutta, M. Patwardhan, P. Deshpande, S. Karande, B. Rai, *Sci Rep* **2023**, *13*, 7347.

[15] S. Harini, P. Deshpande, J. Dutta, B. Rai, in *Proceedings of the 1st International Conference on Water Energy Food and Sustainability (ICoWEFS 2021)* (Eds.: J. R. da Costa Sanches Galvão, P. S. Duque de Brito, F. dos Santos Neves, F. G. da Silva Craveiro, H. de Amorim Almeida, J. O. Correia Vasco, L. M. Pires Neves, R. de Jesus Gomes, S. de Jesus Martins Mourato, V. S. Santos Ribeiro), Springer International Publishing, Cham, **2021**, pp. 187–192.

[16] J. Dutta, M. Chennamkulam Ajith, S. Dutta, U. R. Kadhane, J. Kochupurackal B, B. Rai, *Sci Rep* **2020**, *10*, 15241.

[17] J. Dutta, M. Patwardhan, P. Deshpande, B. Rai, **n.d.**

[18] J. Dutta, P. DESHPANDE, B. Rai, *System and Method for Managing Ripening Conditions of Climacteric Fruits*, **2020**, US20200281220A1.

[19] B. Rai, J. Dutta, P. DESHPANDE, S. B. KAUSLEY, S. S. Karande, M. S. PATWARDHAN, S. M. DESHMUKH, *System and Method for Monitoring and Quality Evaluation of Perishable Food Items*, **2022**, US11488017B2.

[20] A. Gholami, Z. Yao, S. Kim, C. Hooper, M. W. Mahoney, K. Keutzer, *IEEE Micro* **2024**, *44*, 33.

[21] S. Liu, R. M. Radway, X. Wang, J. Kwon, C. Trippel, P. Levis, S. Mitra, H.-S. P. Wong, in *2024 IEEE International Electron Devices Meeting (IEDM)*, IEEE, San Francisco, CA, USA, **2024**, pp. 1–4.

[22] S. Aleksic, in *2017 14th International Conference on Telecommunications (ConTEL)*, **2017**, pp. 41–46.

[23] C. Haffner, W. Heni, Y. Fedoryshyn, J. Niegemann, A. Melikyan, D. L. Elder, B. Baeuerle, Y. Salamin, A. Josten, U. Koch, C. Hoessbacher, F. Ducry, L. Juchli, A. Emboras, D. Hillerkuss, M. Kohl, L. R. Dalton, C. Hafner, J. Leuthold, *Nature Photon* **2015**, *9*, 525.

[24] C. Kachris, I. Tomkos, *IEEE Communications Surveys & Tutorials* **2012**, *14*, 1021.

[25] D. Tsiokos, G. T. Kanellos, in *Optical Interconnects for Data Centers* (Eds.: T. Tekin, R. Pitwon, A. Håkansson, N. Pleros), Woodhead Publishing, **2017**, pp. 43–73.

[26] C. A. Thraskias, E. N. Lallas, N. Neumann, L. Schares, B. J. Offrein, R. Henker, D. Plettemeier, F. Ellinger, J. Leuthold, I. Tomkos, *IEEE Communications Surveys & Tutorials* **2018**, *20*, 2758.

[27] S. S. Gill, H. Wu, P. Patros, C. Ottaviani, P. Arora, V. C. Pujol, D. Haunschild, A. K. Parlikad, O. Cetinkaya, H. Lutfiyya, V. Stankovski, R. Li, Y. Ding, J. Qadir, A. Abraham, S. K. Ghosh, H. H. Song, R. Sakellariou, O. Rana, J. J. P. C. Rodrigues, S. S. Kanhere, S. Dustdar, S. Uhlig, K. Ramamohanarao, R. Buyya, *Telematics and Informatics Reports* **2024**, *13*, 100116.

[28] "Memory Is All You Need: An Overview of Compute-in-Memory Architectures for Accelerating Large Language Model Inference," can be found under https://arxiv.org/html/2406.08413v1, **n.d.**

[29]  "Roadmap to neuromorphic computing with emerging technologies | APL Materials | AIP Publishing," can be found under https://pubs.aip.org/aip/apm/article/12/10/109201/3317314/Roadmap-to-neuromorphic-computing-with-emerging, **n.d.**

[30]  A. Sebastian, M. Le Gallo, R. Khaddam-Aljameh, E. Eleftheriou, *Nat. Nanotechnol.* **2020**, *15*, 529.

[31]  Y. Shen, N. C. Harris, S. Skirlo, M. Prabhu, T. Baehr-Jones, M. Hochberg, X. Sun, S. Zhao, H. Larochelle, D. Englund, M. Soljačić, *Nature Photon* **2017**, *11*, 441.

[32]  J. Capmany, I. Gasulla, D. Pérez, *Nature Photon* **2016**, *10*, 6.

[33]  N. C. Harris, G. R. Steinbrecher, M. Prabhu, Y. Lahini, J. Mower, D. Bunandar, C. Chen, F. N. C. Wong, T. Baehr-Jones, M. Hochberg, S. Lloyd, D. Englund, *Nature Photon* **2017**, *11*, 447.

[34]  "Integrated photonics on thin-film lithium niobate," can be found under https://opg.optica.org/aop/fulltext.cfm?uri=aop-13-2-242&id=450625, **n.d.**

[35]  P. Prabhathan, K. V. Sreekanth, J. Teng, J. H. Ko, Y. J. Yoo, H.-H. Jeong, Y. Lee, S. Zhang, T. Cao, C.-C. Popescu, B. Mills, T. Gu, Z. Fang, R. Chen, H. Tong, Y. Wang, Q. He, Y. Lu, Z. Liu, H. Yu, A. Mandal, Y. Cui, A. S. Ansari, V. Bhingardive, M. Kang, C. K. Lai, M. Merklein, M. J. Müller, Y. M. Song, Z. Tian, J. Hu, M. Losurdo, A. Majumdar, X. Miao, X. Chen, B. Gholipour, K. A. Richardson, B. J. Eggleton, M. Wuttig, R. Singh, *iScience* **2023**, *26*, 107946.

[36]  G. Moody, V. J. Sorger, D. J. Blumenthal, P. W. Juodawlkis, W. Loh, C. Sorace-Agaskar, A. E. Jones, K. C. Balram, J. C. F. Matthews, A. Laing, M. Davanco, L. Chang, J. E. Bowers, N. Quack, C. Galland, I. Aharonovich, M. A. Wolff, C. Schuck, N. Sinclair, M. Lončar, T. Komljenovic, D. Weld, S. Mookherjea, S. Buckley, M. Radulaski, S. Reitzenstein, B. Pingault,

B. Machielse, D. Mukhopadhyay, A. Akimov, A. Zheltikov, G. S. Agarwal, K. Srinivasan, J. Lu, H. X. Tang, W. Jiang, T. P. McKenna, A. H. Safavi-Naeini, S. Steinhauer, A. W. Elshaari, V. Zwiller, P. S. Davids, N. Martinez, M. Gehl, J. Chiaverini, K. K. Mehta, J. Romero, N. B. Lingaraju, A. M. Weiner, D. Peace, R. Cernansky, M. Lobino, E. Diamanti, L. T. Vidarte, R. M. Camacho, *J. Phys. Photonics* **2022**, *4*, 012501.

[37] W. Bogaerts, D. Pérez, J. Capmany, D. A. B. Miller, J. Poon, D. Englund, F. Morichetti, A. Melloni, *Nature* **2020**, *586*, 207.

[38] M. Haselman, S. Hauck, *Proceedings of the IEEE* **2010**, *98*, 11.

[39] S. A. Schulz, Rupert. F. Oulton, M. Kenney, A. Alù, I. Staude, A. Bashiri, Z. Fedorova, R. Kolkowski, A. F. Koenderink, X. Xiao, J. Yang, W. J. Peveler, A. W. Clark, G. Perrakis, A. C. Tasolamprou, M. Kafesaki, A. Zaleska, W. Dickson, D. Richards, A. Zayats, H. Ren, Y. Kivshar, S. Maier, X. Chen, M. A. Ansari, Y. Gan, A. Alexeev, T. F. Krauss, A. Di Falco, S. D. Gennaro, T. Santiago-Cruz, I. Brener, M. V. Chekhova, R.-M. Ma, V. V. Vogler-Neuling, H. C. Weigand, Ü.-L. Talts, I. Occhiodori, R. Grange, M. Rahmani, L. Xu, S. M. Kamali, E. Arababi, A. Faraon, A. C. Harwood, S. Vezzoli, R. Sapienza, P. Lalanne, A. Dmitriev, C. Rockstuhl, A. Sprafke, K. Vynck, J. Upham, M. Z. Alam, I. De Leon, R. W. Boyd, W. J. Padilla, J. M. Malof, A. Jana, Z. Yang, R. Colom, Q. Song, P. Genevet, K. Achouri, A. B. Evlyukhin, U. Lemmer, I. Fernandez-Corbaton, *Appl. Phys. Lett.* **2024**, *124*, 260701.

[40] J. Geler-Kremer, F. Eltes, P. Stark, D. Stark, D. Caimi, H. Siegwart, B. Jan Offrein, J. Fompeyrine, S. Abel, *Nat. Photon.* **2022**, *16*, 491.

[41] K. Gao, K. Du, S. Tian, H. Wang, L. Zhang, Y. Guo, B. Luo, W. Zhang, T. Mei, *Advanced Functional Materials* **2021**, *31*, 2103327.

[42]  P. Moitra, Y. Wang, X. Liang, L. Lu, A. Poh, T. W. W. Mass, R. E. Simpson, A. I. Kuznetsov, R. Paniagua-Dominguez, *Advanced Materials* **2023**, *35*, 2205367.

[43]  X. Li, N. Youngblood, C. Ríos, Z. Cheng, C. D. Wright, W. H. Pernice, H. Bhaskaran, *Optica, OPTICA* **2019**, *6*, 1.

[44]  E. Goi, Q. Zhang, X. Chen, H. Luan, M. Gu, *PhotoniX* **2020**, *1*, 3.

[45]  A. Tsakyridis, M. Moralis-Pegios, G. Giamougiannis, M. Kirtas, N. Passalis, A. Tefas, N. Pleros, *APL Photonics* **2024**, *9*, 011102.

[46]  R. Chen, A. Tang, J. Dutta, V. Tara, J. Ye, Z. Fang, A. Majumdar, **2025**, DOI 10.48550/arXiv.2506.18592.

[47]  R. Chen, V. Tara, J. Dutta, Z. Fang, J. Zheng, A. Majumdar, *JOM* **2024**, *4*, 031202.

[48]  J. Dutta, R. Chen, V. Tara, A. Majumdar, **2025**.

[49]  J. Dutta, A. Ferraro, A. Manna, R. Chen, A. Pane, G. E. Lio, R. Caputo, A. Majumdar, *ACS Photonics* **2025**, DOI 10.1021/acsphotonics.5c00715.

[50]  I. Chakraborty, G. Saha, K. Roy, *Phys. Rev. Appl.* **2019**, *11*, 014063.

[51]  C. Lian, C. Vagionas, T. Alexoudi, N. Pleros, N. Youngblood, C. Ríos, *Nanophotonics* **2022**, *11*, 3823.

[52]  Z. Zhu, G. Di Guglielmo, Q. Cheng, M. Glick, J. Kwon, H. Guan, L. P. Carloni, K. Bergman, *Journal of Lightwave Technology* **2020**, *38*, 2815.

[53]  S. Abdollahramezani, O. Hemmatyar, M. Taghinejad, H. Taghinejad, A. Krasnok, A. A. Eftekhar, C. Teichrib, S. Deshmukh, M. A. El-Sayed, E. Pop, M. Wuttig, A. Alù, W. Cai, A. Adibi, *Nat Commun* **2022**, *13*, 1696.

[54]  R. Matos, N. Pala, *Micromachines (Basel)* **2023**, *14*, 1259.

[55] E.-S. Lee, J. E. Yoo, D. S. Yoon, S. D. Kim, Y. Kim, S. Hwang, D. Kim, H.-C. Jeong, W. T. Kim, H. J. Chang, H. Suh, D.-H. Ko, C. Cho, Y. Choi, D. H. Kim, M.-H. Cho, *Sci Rep* **2020**, *10*, 13673.

[56] R. Xu, S. Taheriniya, A. P. Ovvyan, J. R. Bankwitz, L. McRae, E. Jung, F. Brückerhoff-Plückelmann, I. Bente, F. Lenzini, H. Bhaskaran, W. H. P. Pernice, *Opt. Mater. Express, OME* **2023**, *13*, 3553.

[57] Z. Fang, R. Chen, J. Zheng, A. Majumdar, *IEEE J. Select. Topics Quantum Electron.* **2022**, *28*, 1.

[58] J. Feldmann, M. Stegmaier, N. Gruhler, C. Ríos, H. Bhaskaran, C. D. Wright, W. H. P. Pernice, *Nat Commun* **2017**, *8*, 1256.

[59] S. Blundell, T. W. Radford, I. A. Ajia, D. Lawson, X. Yan, M. Banakar, D. J. Thomson, I. Zeimpekis, O. L. Muskens, *ACS Photonics* **2025**, *12*, 1382.

[60] M. De Carlo, F. De Leonardis, R. Soref, V. M. N. Passaro, *Journal of Lightwave Technology* **2025**, *43*, 3429.

[61] D. Pérez, I. Gasulla, L. Crudgington, D. J. Thomson, A. Z. Khokhar, K. Li, W. Cao, G. Z. Mashanovich, J. Capmany, *Nat Commun* **2017**, *8*, 636.

[62] W. Zhang, J. Yao, *Nat Commun* **2020**, *11*, 406.

[63] G. T. Reed, G. Mashanovich, F. Y. Gardes, D. J. Thomson, *Nature Photon* **2010**, *4*, 518.

[64] C. Wang, M. Zhang, X. Chen, M. Bertrand, A. Shams-Ansari, S. Chandrasekhar, P. Winzer, M. Lončar, *Nature* **2018**, *562*, 101.

[65] A. Melikyan, L. Alloatti, A. Muslija, D. Hillerkuss, P. C. Schindler, J. Li, R. Palmer, D. Korn, S. Muehlbrandt, D. Van Thourhout, B. Chen, R. Dinu, M. Sommer, C. Koos, M. Kohl, W. Freude, J. Leuthold, *Nature Photon* **2014**, *8*, 229.

[66]    A. Prencipe, K. Gallo, *IEEE Journal of Quantum Electronics* **2023**, *59*, 1.

[67]    R. Chen, Z. Fang, F. Miller, H. Rarick, J. E. Fröch, A. Majumdar, *ACS Photonics* **2022**, *9*, 3181.

[68]    K. Aryana, C. C. Popescu, H. Sun, K. Aryana, H. J. Kim, M. Julian, M. R. Islam, C. A. Ríos Ocampo, T. Gu, J. Hu, P. E. Hopkins, *Advanced Materials* **2025**, *37*, 2414031.

[69]    S. Blundell, T. W. Radford, I. A. Ajia, D. Lawson, X. Yan, M. Banakar, D. J. Thomson, I. Zeimpekis, O. L. Muskens, *ACS Photonics* **2025**, *12*, 1382.

[70]    F. M. Schenk, T. Zellweger, D. Kumaar, D. Bošković, S. Wintersteller, P. Solokha, S. De Negri, A. Emboras, V. Wood, M. Yarema, *ACS Nano* **2023**, *18*, 1063.

[71]    J. Meng, Y. Gui, B. M. Nouri, X. Ma, Y. Zhang, C.-C. Popescu, M. Kang, M. Miscuglio, N. Peserico, K. Richardson, J. Hu, H. Dalir, V. J. Sorger, *Light Sci Appl* **2023**, *12*, 189.

[72]    R. Chen, Z. Fang, C. Perez, F. Miller, K. Kumari, A. Saxena, J. Zheng, S. J. Geiger, K. E. Goodson, A. Majumdar, *Nat Commun* **2023**, *14*, 3465.

[73]    Z. Fang, R. Chen, J. Zheng, A. I. Khan, K. M. Neilson, S. J. Geiger, D. M. Callahan, M. G. Moebius, A. Saxena, M. E. Chen, C. Rios, J. Hu, E. Pop, A. Majumdar, *Nat. Nanotechnol.* **2022**, *17*, 842.

[74]    M. Wei, K. Xu, B. Tang, J. Li, Y. Yun, P. Zhang, Y. Wu, K. Bao, K. Lei, Z. Chen, H. Ma, C. Sun, R. Liu, M. Li, L. Li, H. Lin, *Nat Commun* **2024**, *15*, 2786.

[75]    C. C. Popescu, K. Aryana, B. Mills, T. W. Lee, L. Martin-Monier, L. Ranno, J. X. B. Sia, K. P. Dao, H.-B. Bae, V. Liberman, S. A. Vitale, M. Kang, K. A. Richardson, C. A. Ríos Ocampo, D. Calahan, Y. Zhang, W. M. Humphreys, H. J. Kim, T. Gu, J. Hu, *Advanced Optical Materials* **2025**, *13*, 2402751.

[76]    "(PDF) A New Family of Ultralow Loss Reversible Phase-Change Materials for Photonic Integrated Circuits: Sb 2 S 3 and Sb 2 Se 3," can be found under https://www.researchgate.net/publication/342814269_A_New_Family_of_Ultralow_Loss_Reversible_Phase-Change_Materials_for_Photonic_Integrated_Circuits_Sb_2_S_3_and_Sb_2_Se_3, **n.d.**

[77]    R. Chen, V. Tara, M. Choi, J. Dutta, J. Sim, J. Ye, Z. Fang, J. Zheng, A. Majumdar, *npj Nanophoton.* **2024**, *1*, 1.

[78]    X. Yang, L. Lu, Y. Li, Y. Wu, Z. Li, J. Chen, L. Zhou, *Advanced Functional Materials* **2023**, *33*, 2304601.

[79]    Z. Fang, B. Mills, R. Chen, J. Zhang, P. Xu, J. Hu, A. Majumdar, *Nano Lett.* **2024**, *24*, 97.

[80]    P. Xu, J. Zheng, J. K. Doylend, A. Majumdar, *ACS Photonics* **2019**, *6*, 553.

[81]    M.-K. Song, J.-H. Kang, X. Zhang, W. Ji, A. Ascoli, I. Messaris, A. S. Demirkol, B. Dong, S. Aggarwal, W. Wan, S.-M. Hong, S. G. Cardwell, I. Boybat, J. Seo, J.-S. Lee, M. Lanza, H. Yeon, M. Onen, J. Li, B. Yildiz, J. A. Del Alamo, S. Kim, S. Choi, G. Milano, C. Ricciardi, L. Alff, Y. Chai, Z. Wang, H. Bhaskaran, M. C. Hersam, D. Strukov, H.-S. P. Wong, I. Valov, B. Gao, H. Wu, R. Tetzlaff, A. Sebastian, W. Lu, L. Chua, J. J. Yang, J. Kim, *ACS Nano* **2023**, *17*, 11994.

[82]    C. Wu, H. Yu, H. Li, X. Zhang, I. Takeuchi, M. Li, *ACS Photonics* **2019**, *6*, 87.

[83]    X. Yang, S. Ran, Z. Li, L. Lu, Y. Li, N. P. Wai, M. Zhang, G.-Q. Lo, J. Chen, L. Zhou, *PhotoniX* **2025**, *6*, 16.

[84]    J. Xia, T. Wang, Z. Wang, J. Gong, Y. Dong, R. Yang, X. Miao, *ACS Photonics* **2024**, *11*, 723.

[85]  C. Ríos, Q. Du, Y. Zhang, C.-C. Popescu, M. Y. Shalaginov, P. Miller, C. Roberts, M. Kang, K. A. Richardson, T. Gu, S. A. Vitale, J. Hu, *PhotoniX* **2022**, *3*, 26.

[86]  J. Liu, C. Zhang, J. Chen, W. Zeng, P. Hu, Q. Liu, *Appl. Opt., AO* **2025**, *64*, 100.

[87]  R. Marchetti, C. Lacava, L. Carroll, K. Gradkowski, P. Minzioni, *Photon. Res.* **2019**, *7*, 201.

[88]  J. Liu, C. Zhang, J. Chen, W. Zeng, P. Hu, Q. Liu, *Appl. Opt., AO* **2025**, *64*, 100.

[89]  G. Kurczveil, P. Pintus, M. J. R. Heck, J. D. Peters, J. E. Bowers, *IEEE Photonics J.* **2013**, *5*, 6600410.

[90]  C. C. Popescu, K. Aryana, B. Mills, T. W. Lee, L. Martin-Monier, L. Ranno, J. X. B. Sia, K. P. Dao, H.-B. Bae, V. Liberman, S. A. Vitale, M. Kang, K. A. Richardson, C. A. Ríos Ocampo, D. Calahan, Y. Zhang, W. M. Humphreys, H. J. Kim, T. Gu, J. Hu, *Advanced Optical Materials* **2025**, *13*, 2402751.

[91]  G. Li, A. V. Krishnamoorthy, I. Shubin, J. Yao, Y. Luo, H. Thacker, X. Zheng, K. Raj, J. E. Cunningham, *IEEE Journal of Selected Topics in Quantum Electronics* **2013**, *19*, 95.

[92]  D. Hall, D. Liang, *High-Index-Contrast Waveguide*, **2008**, US20080267239A1.

[93]  J. Zheng, Z. Fang, C. Wu, S. Zhu, P. Xu, J. K. Doylend, S. Deshmukh, E. Pop, S. Dunham, M. Li, A. Majumdar, *Advanced Materials* **2020**, *32*, 2001218.

[94]  W. Bogaerts, P. De Heyn, T. Van Vaerenbergh, K. De Vos, S. Kumar Selvaraja, T. Claes, P. Dumon, P. Bienstman, D. Van Thourhout, R. Baets, *Laser & Photonics Reviews* **2012**, *6*, 47.

**Supporting Information**

Additional details are provided in the Supplementary Information file, including numerical simulations (Lumerical eigenmode analysis, Finite Difference Time Domain (FDTD) simulation and COMSOL heat-transfer modeling), a step-by-step fabrication flow, electrical and optical micrographs of the fabricated devices, and extended structural characterization (SEM and AFM) across different PCM geometries. Further analyses examine taper-length dependence, island-number scaling, and extended endurance behavior, providing additional perspective on the performance trade-offs of PCM–PIC architectures. Supporting Information is available from the Wiley Online Library or from the author.

**Table of Contents**

Tapered and segmented islands of the phase-change material (PCM) antimony selenide ($Sb_2Se_3$) integrated on silicon microring resonators enable ultra-low-loss, high-endurance, non-volatile photonic tuning. The nanostructured PCM design suppresses interfacial scattering and thermal stress, achieving record endurance over 100 million electrical switching cycles with ultra-low loss, low-voltage actuation and strong optical contrast, establishing a CMOS-compatible platform for reliable, reconfigurable photonic circuits.

**Title: Increased endurance of nonvolatile photonics enabled by nanostructured phase-change materials**

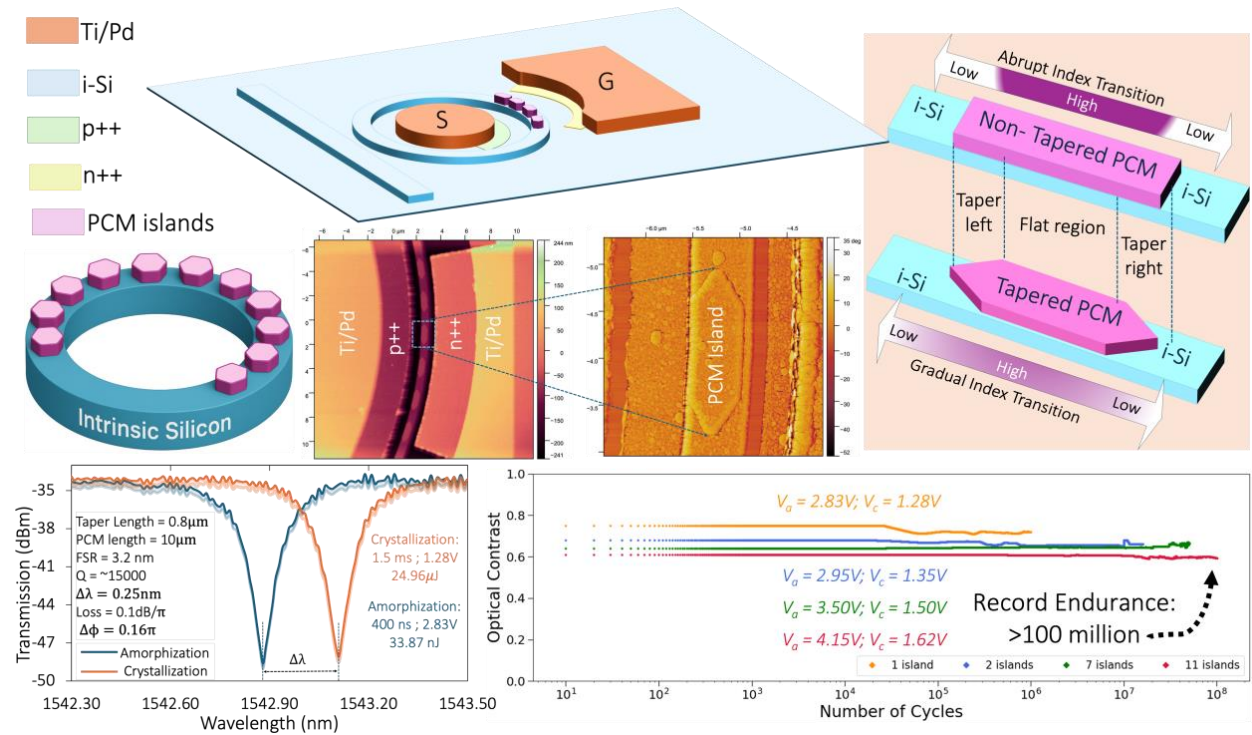